\documentclass[fleqn,usenatbib]{mnras}

\usepackage{newtxtext,newtxmath}

\usepackage[T1]{fontenc}

\DeclareRobustCommand{\VAN}[3]{#2}
\let\VANthebibliography\thebibliography
\def\thebibliography{\DeclareRobustCommand{\VAN}[3]{##3}\VANthebibliography}

\usepackage{graphicx}	
\usepackage{amsmath}	

\title[Shocked emission outside the AGN ionization axis]{Chemical abundances in Seyfert galaxies -- V. The discovery of shocked emission outside the AGN ionization axis}
\author[R. A. Riffel et al.]{
R. A. Riffel,$^{1}$\thanks{E-mail: rogemar@ufsm.br}
O. L. Dors,$^{2}$ 
M. Armah,$^{2}$
T. Storchi-Bergmann,$^{3}$,
A. Feltre,$^{4}$
\newauthor
G. F. H\"agele,$^{5,6}$
M. V. Cardaci,$^{5,6}$
D. Ruschel-Dutra,$^{7}$
A. C Krabbe,$^{2}$
E. P\'erez-Montero,$^{8}$
\newauthor
N. L. Zakamska,$^{9}$
I. C. Freitas$^{10}$
\\
$^{1}$Departamento de F\'isica, Centro de Ci\^encias Naturais e Exatas, Universidade Federal de Santa Maria, 97105-900, Santa Maria, RS, Brazil\\
$^{2}$Universidade do Vale do Para\'iba, Av. Shishima Hifumi, 2911, Cep
12244-000, S\~ao Jos\'e dos Campos, SP, Brazil \\
$^{3}$Instituto de F\'isica, Universidade Federal do Rio Grande do Sul,  CP 15051, Porto Alegre, RS, 91501-970, Brazil\\
$^{4}$ INAF - Osservatorio di Astrofisica e Scienza dello Spazio di Bologna, Via P. Gobetti 93/3, 40129 Bologna, Italy\\
$^{5}$ Instituto de Astrof\'{i}sica de La Plata (CONICET-UNLP), Argentina \\
$^{6}$ Facultad de Ciencias Astron\'{o}micas y Geof\'{i}sicas, Universidad Nacional de La Plata, Paseo del Bosque s/n, 1900 La Plata, Argentina \\
$^{7}$Departamento de F\'isica, Universidade Federal de Santa Catarina, P.O. Box 476, 88040-900, Florian\'opolis, SC, Brazil \\
$^{8}$ Instituto de Astrof{\'i}sica de Andaluc{\'i}a, Camino Bajo de Hu{\'e}tor s/n, Aptdo. 3004, E18080-Granada, Spain. \\
$^{9}$Department of Physics \& Astronomy, Johns Hopkins University, Bloomberg Center, 3400 N. Charles St, Baltimore, MD 21218, USA\\
$^{10}$Col\'egio Polit\'ecnico, Universidade Federal de Santa Maria, Santa Maria, 97105-900 RS, Brazil
}

\date{Accepted XXX. Received YYY; in original form ZZZ}

\pubyear{2020}

\begin{document}
\label{firstpage}
\pagerange{\pageref{firstpage}--\pageref{lastpage}}
\maketitle

\begin{abstract}
We present maps for the electron temperature in the inner kpc of three luminous Seyfert galaxies: Mrk\,79, Mrk\,348, and Mrk\,607 obtained from Gemini GMOS-IFU observations at spatial resolutions of $\sim$110--280 pc.
We study the distributions of electron temperature in active galaxies and find temperatures varying in the range from $\sim$8\,000 to $\gtrsim30\,000 $K. Shocks due to gas outflows play an important role in the observed temperature distributions of Mrk\,79 and Mrk\,348, while standard photoionization models  reproduce the derived temperature values for Mrk\,607. In Mrk\,79 and Mrk\,348, we find direct evidence for shock-ionization with overall orientation orthogonal to the ionization axis, where shocks can be easily observed as the AGN radiation field is shielded by the nuclear dusty torus. This also indicates that even when the ionization cones are narrow, the shocks can be much wider-angle.
\end{abstract}




\section{Introduction}

\begin{figure*}
\includegraphics[width=0.73\textwidth]{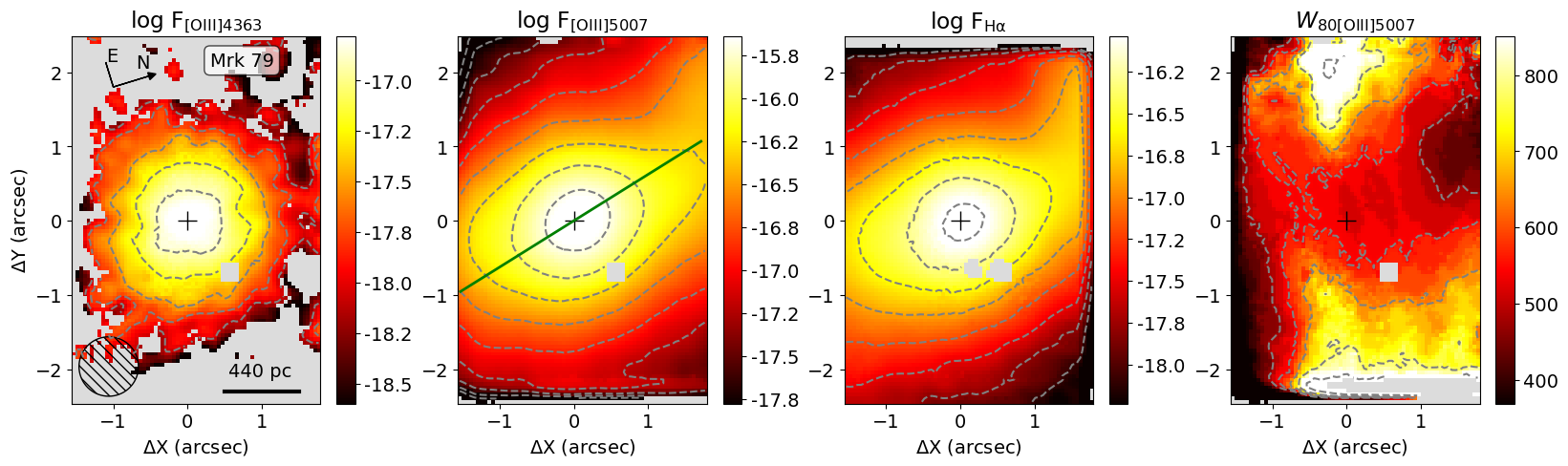}
\includegraphics[width=0.73\textwidth]{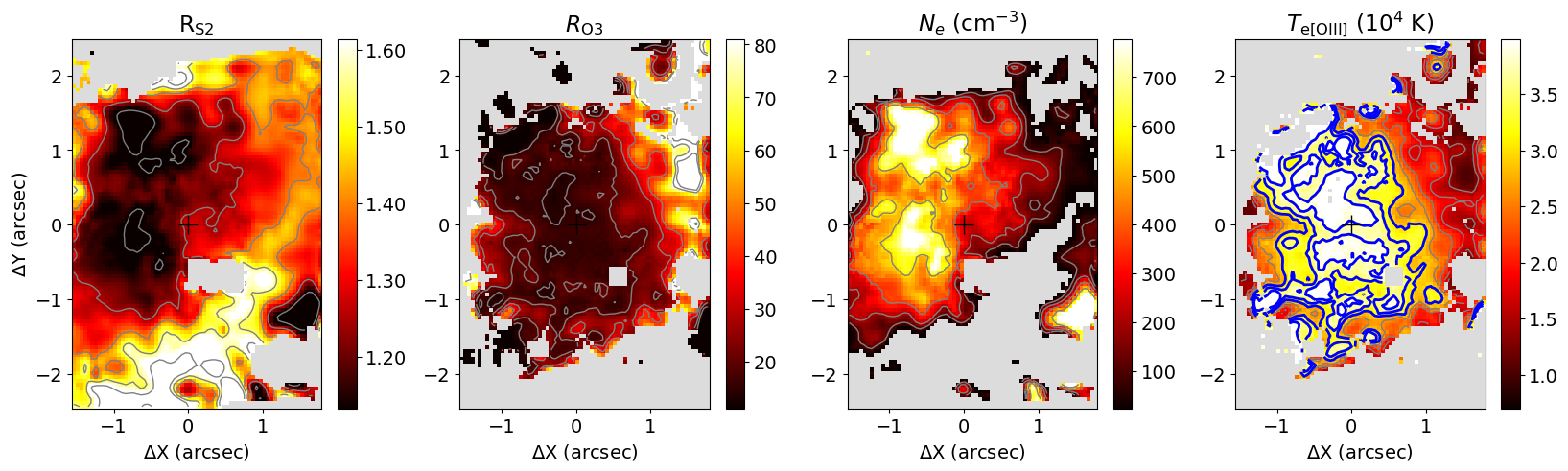}
\caption{Maps for Mrk\,79, which has a bolometric luminosity of log\,$L_{\rm bol}$/(erg\,s$^{-1}$)$=45.0$ and a distance of $d=91.6$\,Mpc \citep[see ][]{Freitas18}. Top row: Emission-line flux distributions and $W_{\rm 80}$ map for the [\ion{O}{iii}]5007 emission line. Bottom row: flux line ratios, $N_e$ and $T_{\rm e[O III]}$ maps. The central crosses mark the position of the continuum peak, the circle shows the seeing disc, the spatial scale and orientation are shown in the [\ion{O}{iii}]$\lambda$4363 flux map.  Gray regions correspond to locations where the emission-lines were not detected at with a $\rm S/N \:> \:3$.  The contours show the levels of each map. Blue contours in the $T_{\rm e[O III]}$ map correspond to values larger than 30\,000 K, where the uncertainties using $R_{\rm O3}$ to derive $T_{\rm e[O III]}$ are large. The green line on the [\ion{O}{iii}] flux maps shows the orientation of the AGN ionization axis, as obtained from HST images by \citet{schmitt03}. }
\label{fig:m79}
\end{figure*}

\begin{figure*}
\includegraphics[width=0.73\textwidth]{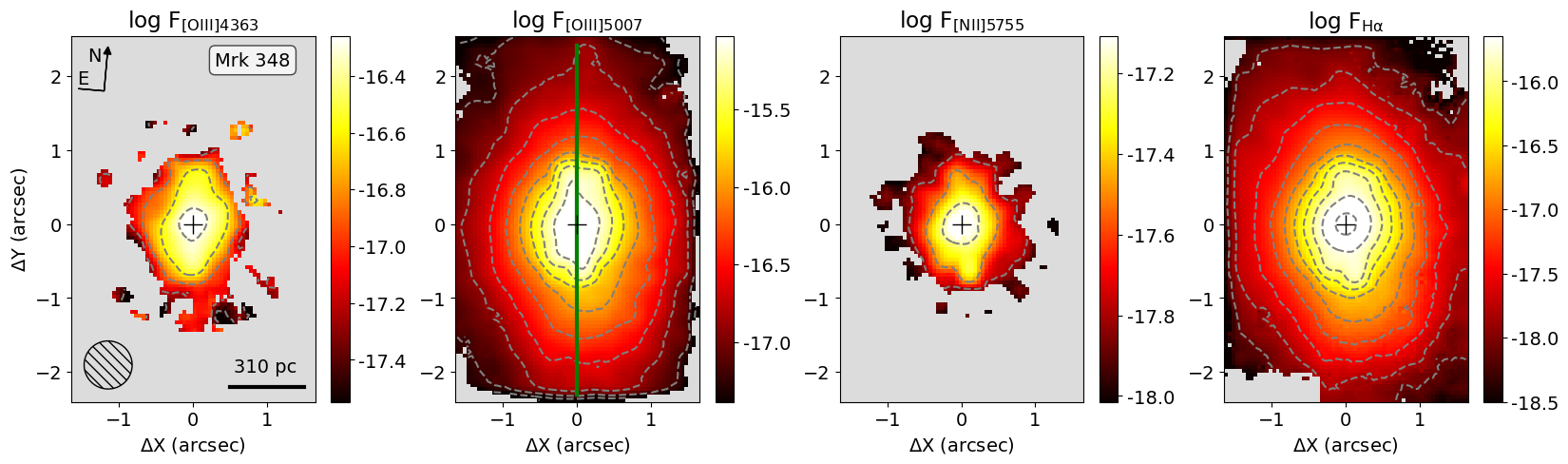}
\includegraphics[width=0.73\textwidth]{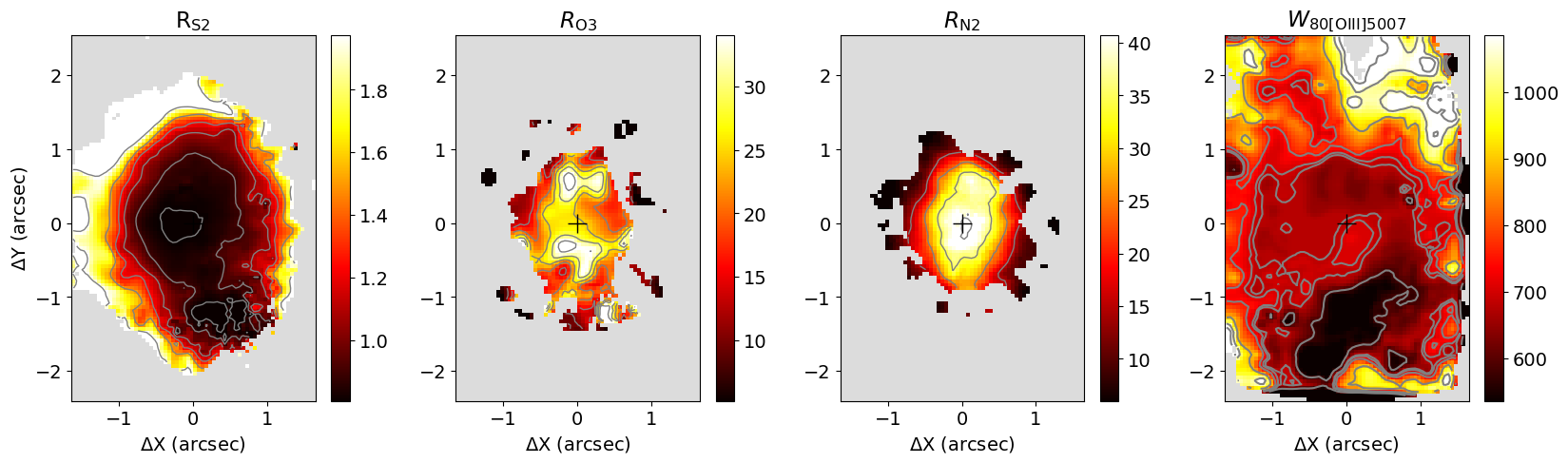}
\includegraphics[width=0.55\textwidth]{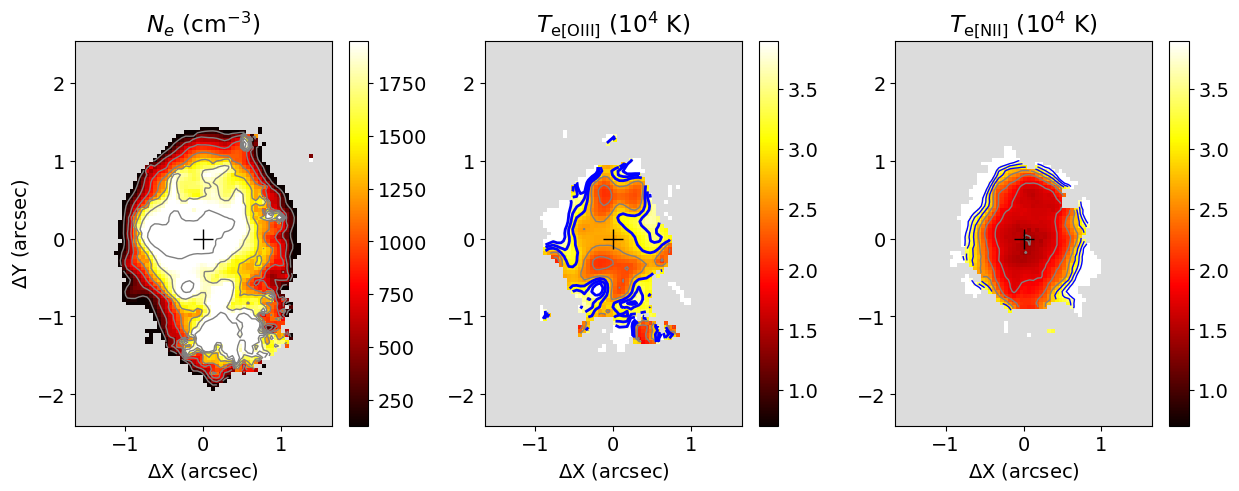}
\caption{Maps for Mrk\,348 (log\,$L_{\rm bol}$/(erg\,s$^{-1}$)$=45.3$, $d=63.9$\,Mpc).  Emission-line flux distributions (top row), flux line ratios and $W_{\rm 80[OIII]}$  (middle row), and $N_e$, $T_{\rm e[O III]}$ and $T_{\rm e[N II]}$ maps (bottom row). The labels are the same as for Fig.~\ref{fig:m79}. }
\label{fig:m348}
\end{figure*}

\begin{figure*}
\includegraphics[width=0.7\textwidth]{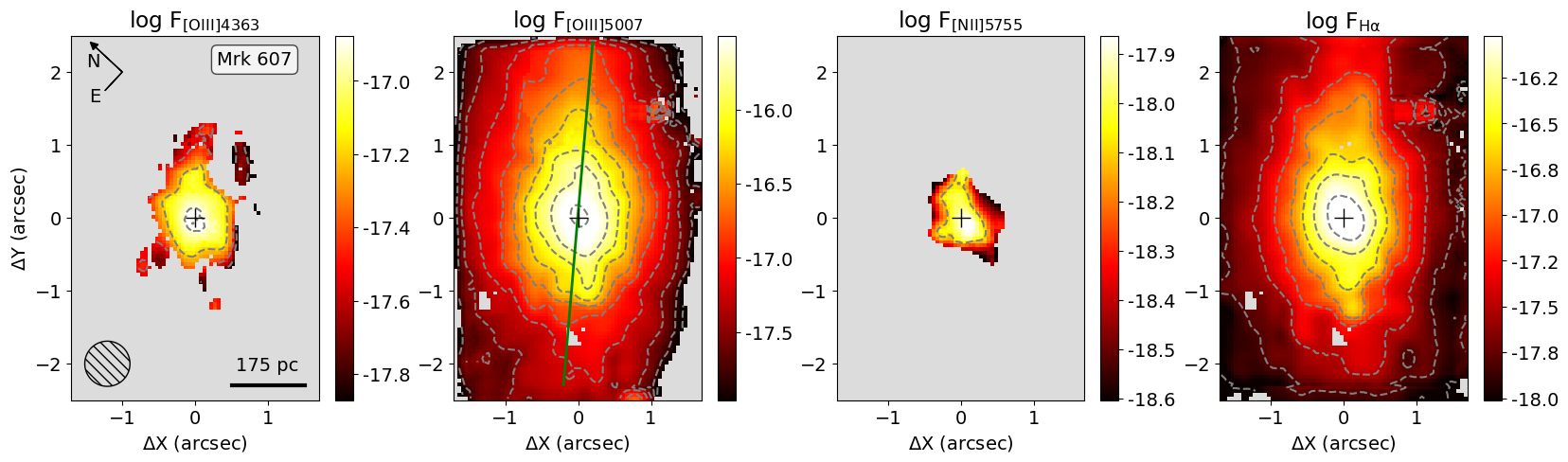}
\includegraphics[width=0.7\textwidth]{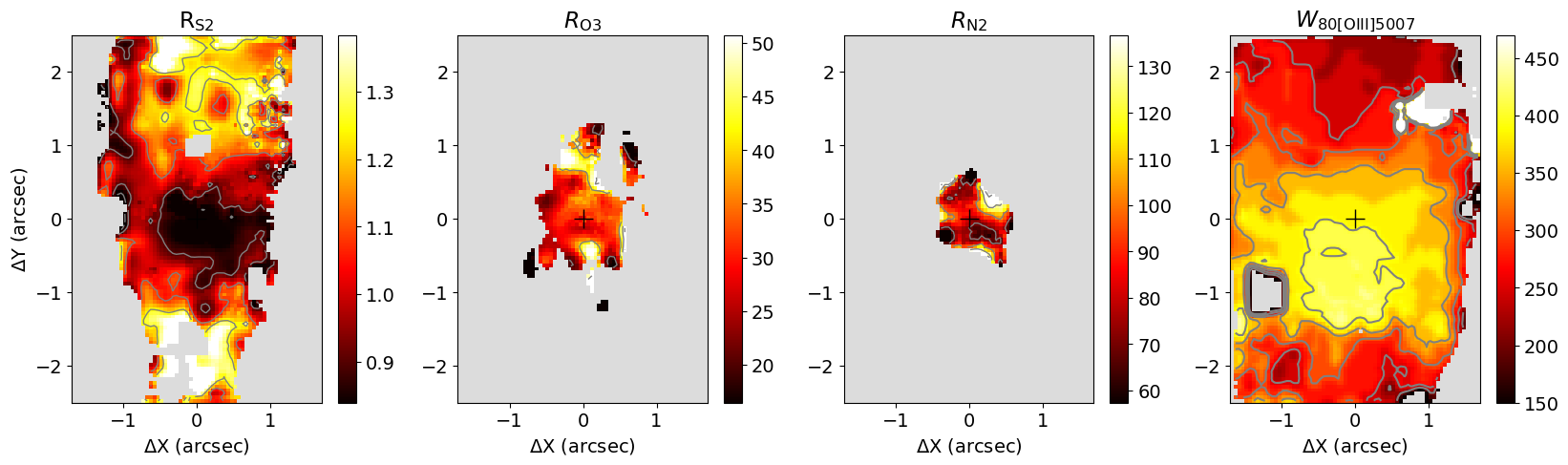}
\includegraphics[width=0.52\textwidth]{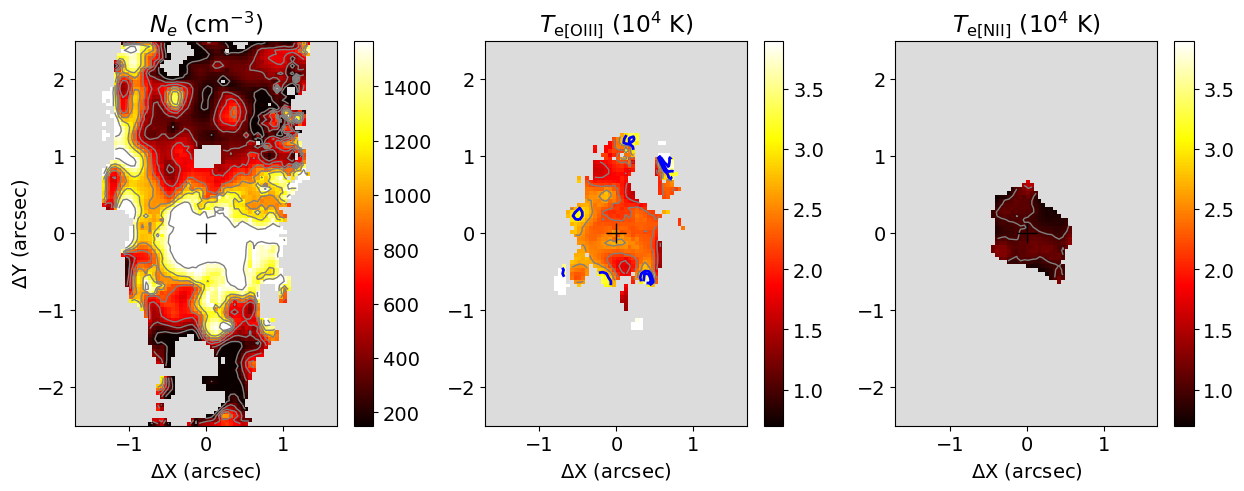}
\caption{Same as Fig.~\ref{fig:m348} but for Mrk\,607 (log\,$L_{\rm bol}$/(erg\,s$^{-1}$)$=43.4$, $d=36.1$\,Mpc). }
\label{fig:m607}
\end{figure*}

Active Galactic Nuclei (AGN) present in their spectra strong emission lines whose  relative intensities
can be used to estimate or characterize the physical and chemical properties of the gas phase in these objects, such as the nature and strength of the interstellar radiation fields, chemical abundance, local temperature, and gas density.
The high luminosity of the AGN continuum and these emission lines have made AGN essential in the studies of chemical evolution of galaxies across the Hubble time. Photoionization models of the Narrow Line Region (NLR) have often been used for determining chemical abundances of heavy elements in AGN host galaxies. The pioneering work by \citet{ferland83}, who used the first version of the photoionization \textsc{Cloudy} code, showed that NLR exhibit metallicities in the range $0.1 \: \la \: (Z/{\rm Z_{\odot}}) \: \la \: 1.0$. Thereafter, several studies exploiting photoionization models have been carried out aiming to estimate the metallicity of AGN host galaxies  in the local universe  \citep[e.g.][]{thaisa98, feltre16, castro17} and at high redshifts \citep[e.g.][]{dors18, mignoli19, guo20}. 

In spite of the wide use of photoionization models in the estimation of gas metallicities, their application in deriving elemental abundances has several limitations. Firstly, it is necessary to assume an incident Spectral Energy Distribution as one of the input parameters of the models, which is generally represented by the power law component of the non-thermal X-ray radiation with continuum between 2 keV and 2\,500\,{\AA} with a spectral index $\alpha_{ox}$ \citep{marchese12,dors19}. Detailed photoionization models used to derive O and N abundances  \citep{dors17} and bayesian-like approach \citep{enrique19} have predicted
$\alpha_{ox}$ to be higher than $-1.2$ for Seyfert~2 nuclei. 
Conversely, measurements of $\alpha_{ox}$ by \citet{miller11} indicate that most AGN have $\alpha_{ox}$ of the order of $-1.4$ and even lower values ($\sim-2.0$)
can be derived for these objects. This indicates that an extra physical process is missing in the models  \citep[probably shocks, e.g.,][]{contini19}. Additionally, the use of emission line intensities observed
in a limited spectral range can produce a  degeneracy in the models \citep[e.g.][]{davies14}
and, consequently, uncertainties in the chemical abundance values.

Direct determination of abundances, known as the $T_{\rm e}$-method which is based on the observational determination of the electron temperature ($T_{\rm e}$), yields more reliable abundance estimations because it circumvents the aforementioned problems of photoionization models. Recently, \citet{dors20} presented a new methodology for the $T_{\rm e}$-method  to be applied to NLRs of Seyfert 2 nuclei, introducing a new relation between temperatures of the  low ($t_{2}$) and high ($t_{3}$) ionization 
gas zones derived from photoionization models. This method produces O/H abundances slightly lower (about 0.2 dex) than those derived from detailed photoionization models. However, previous studies have questioned the use of the $T_{\rm e}$-method in AGN.
For example, \citet{stasisnka84} argued that in the NLR the intensity of the [\ion{O}{iii}]($\lambda$4949+$\lambda$5007)/$\lambda$4363 line ratio, used to derive $t_{3}$ \citep[e.g.][]{hagele08},  is enhanced by emissions from clouds with high gas density 
($N_{\rm e} \: \ga \: 10^{5} \: \mathrm{cm^{-3}}$), which precludes any direct determination of abundances based on the $T_{\rm e}$-method. \citet{nagao01} found evidence that the [\ion{O}{iii}]$\lambda$4363 line
is emitted in denser   ($N_{\rm e} \: \sim \: 10^{5-7} \: \rm cm^{-3}$)
and obscured gas regions than those emitting  [\ion{O}{iii}]$\lambda$5007. 

Shocks from AGN outflows may also affect the ionization structure and $T_{\rm e}$ distribution, particularly in regions where the AGN ionizing photons are shielded by nuclear obscuration \citep{contini01,zakamska14}, where the contribution of photoionization to gas ionization is expected to be lower. Spatially resolved measurements of electron temperature and density are important to constrain the AGN contribution to gas ionization and to infer the gas physical conditions.   Observational data that enable  measurements of  weak auroral lines, from which $T_{\rm e}$ can be estimated, and spatially resolved studies of $T_{\rm e}$ in AGN are seldom  found in the literature \citep[but see][]{revalski18,dahmer19,dagostino19}. 

In this Letter, we present two-dimensional (2D) maps of electron temperature and electron density in the NLR, in the case in one Seyfert~1 (Mrk\,79) and two Seyfert~2  (Mrk\,348 and Mrk\,607) galaxies, obtained from integral field spectroscopy. 
A thorough analysis of $T_{\mathrm{e}}$ and $N_{\mathrm{e}}$ could 
be performed with the use of 2D spectroscopy at spatial resolutions of a few hundred parsecs in these 3 AGN hosts. In Sect.~\ref{datasec}, we describe the data and techniques to derive $T_{\mathrm{e}}$ and $N_{\mathrm{e}}$. In Sect.~\ref{resulsec} we present and discuss the results,  while the conclusions  are given in Sect.~\ref{concsec}.

\section{Methodology}
\label{datasec}

We use the Gemini Multi-Object Spectrograph \citep[GMOS,][]{gmos} Integral Field Unit (IFU) data to map the $T_{\mathrm{e}}$ and $N_{\mathrm{e}}$ in the inner few hundred parsecs of the nearby luminous Seyfert galaxies Mrk\,79 (Sy 1, SBb), Mrk\,348 (Sy 2, SA(s)0/a) and Mrk\,607 (Sy 2,  Sa). These galaxies were selected from the sample of \citet{Freitas18} because they have [\ion{O}{iii}]$\lambda$4363 extended emission.
The GMOS data covers the spectral range from 4300 to 7100\,\AA\, with a velocity resolution of $\sim$90\,km\,s$^{-1}$ (FWHM) and spatial resolutions of 280$\pm$30 (Mrk\,79), 190$\pm$25 (Mrk\,348), and 110$\pm$14 pc (Mrk\,607).
The data reduction followed the standard procedures using the {\sc gemini.iraf} package as described by \citet{Freitas18}. 
We use the {\sc ifscube} python package \citep{ifscube} to fit the emission-line profiles and obtain the emission-line flux distributions. 
We allow the fit of up to three Gaussian components per emission line
and the line fluxes are obtained by the sum of the fluxes of the individual components (see Suppl.  Mat.).
We derive the electron temperature using two sets of auroral/nebular line intensity ratios: $R_{\rm O3}$ = $([\ion{O}{iii}]\lambda\lambda 4959,5007/\lambda4363)$ and $R_{N2}$ = $([\ion{N}{ii}]\lambda\lambda 6548,6584/\lambda5755)$, using \citep{hagele08}:

\begin{equation}
 \label{eqt3}
 \frac{T_{\rm e[O III]}}{\rm 10^4 K} = 0.8254-0.0002415 R_{\rm O3}+\frac{47.77}{R_{\rm O3}}
 \end{equation}
 and
 \begin{equation}
\label{eqtn2}
\frac{T_{\rm e[N II]}}{\rm 10^4 K}=0.537+0.000253 \times R_{\rm N2}+\frac{42.13}{R_{\rm N2}}.
\end{equation}
 The [\ion{N}{ii}]$\lambda$\,5755 emission line is not detected for Mrk\,79 and, thus we are unable to  calculate the $T_{\rm e[N II]}$  for this galaxy. 

We estimate the electron density ($N_{\mathrm {e}}$) from
the $R_{S2}$ = [\ion{S}{ii}]$\lambda 6716/\lambda 6731$ emission-line intensity ratio using the {\sc PyNeb} routine \citep{luridiana15},  assuming the $t_{3}$ values obtained for each spaxel. For spaxels with no measurements of $t_{3}$, we use the mean $t_{3}$ value for each galaxy. 

\section{Results \& Discussion}
\label{resulsec}

Figures \ref{fig:m79}, \ref{fig:m348}, and \ref{fig:m607} show emission-line flux distributions, flux line ratios, electron density, and electron temperature maps for Mrk\,79, Mrk\,348, and Mrk\,607, respectively. The emission line ratios are  corrected for dust extinction, as estimated from the H$\alpha$ and H$\beta$ fluxes (see Suppl. Mat.). The flux distributions for [\ion{O}{iii}]$\lambda$4363, [\ion{O}{iii}]$\lambda$5007, [\ion{N}{ii}]$\lambda$5755, and H$\alpha$ emission lines are shown in the first row of each figure. There is no detection of the [\ion{N}{ii}]$\lambda$5755 emission line in the spectra of Mrk\,79 and thus, we do not show its flux map for this galaxy. The $N_{\rm e}$ maps present values in the range $\sim$100--2\,000 cm$^{-3}$, with the highest values seen at the nucleus for all galaxies. The [\ion{Ar}{iv}]$\lambda$4711, $\lambda$4740 emission lines trace denser gas phases
than the [\ion{S}{ii}] lines ratio. The [\ion{Ar}{iv}] emission is not spatially resolved in our sample, but we measure the [\ion{Ar}{iv}]$\lambda$4711/$\lambda$4740 line ratio by integrating the spectra within an 1$^{\prime\prime}\times$1$^{\prime\prime}$ centred at the nucleus. We obtain ratios of $0.64\pm0.20$,  $0.44\pm0.15$ and $0.25\pm0.20$, which correspond to densities of $\sim$15\,200, 31\,700  and 80\,800 cm$^{-3}$ for Mrk\,79, Mrk\,348 and Mrk\,607, respectively.

\begin{figure}
{\centering
\includegraphics[width=0.45\textwidth]{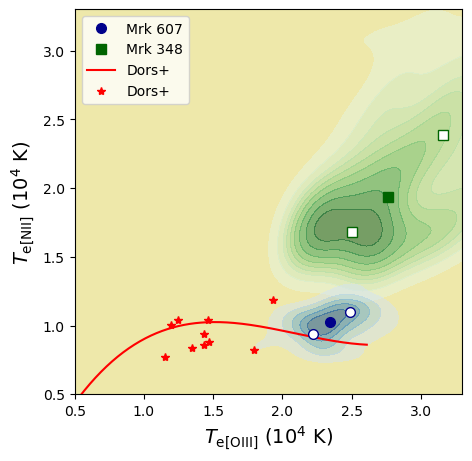}
}
\caption{$T_{\rm e[O III]}$ vs. $T_{\rm e[N II]}$. The green and blue  density curves are for Mrk\,348 and Mrk\,607 respectively.The open squares and open circles show the 25th and 75th percentile $T_{\rm e}$ values for Mrk\,348 and Mrk\,607, respectively. The filled squares and circles show the median $T_{\rm e}$ values.  The red stars are the AGN compilation of direct estimates of temperatures from \citet{dors17,dors20}  and the red curve corresponds to  predictions of their AGN photoionization models.}
\label{golden}
\end{figure}

The [\ion{O}{iii}]$\lambda$4363 and [\ion{N}{ii}]$\lambda$5755 flux distributions are spatially resolved (see Suppl. Mat.) in our sample.  Thus, we use the fluxes of these lines to compute spatially resolved maps for $R_{\mathrm{O3}}$ and $R_{\mathrm{N2}}$. The [\ion{S}{ii}] emission spreads over most of the GMOS field of view for all galaxies, allowing us to construct  $R_{S2}$ maps. We show these maps in Figs.~\ref{fig:m79}, \ref{fig:m348} and \ref{fig:m607}.

The bottom panels of Figs.~\ref{fig:m79}, \ref{fig:m348}, and \ref{fig:m607} show the $N_{\rm e}$ and $T_{\rm e}$ maps for Mrk\,79, Mrk\,348, and Mrk\,607, respectively. These properties were computed from the $R_{S2}$, $R_{\rm O3}$, $R_{N2}$ following the procedure described in Sec.~\ref{datasec}.
 In all galaxies, we find the highest values of $R_{\mathrm{O3}}$ along the AGN photo-ionization structure traced by the [\ion{O}{iii}]$\lambda5007$ emission \citep{schmitt03}. Otherwise, smaller $R_{\mathrm{O3}}$ values are derived mostly away from the AGN ionization axis, indicating higher temperatures at these locations. This indicates that, besides AGN photoionization, an additional process may be producing the observed emission.  There are some differences in the  $T_{\rm e}$  distributions observed in the three objects.   For Mrk\,348 and  Mrk\,79 very high temperature values ($\gtrsim 30\,000$ K) are observed in several spaxels. 
Most of the ionized gas emission in the inner kpc of these galaxies is produced in outflows with velocities of up to $\rm 200 \: km\: s^{-1}$ \citep{Freitas18,rogemar_mrk79}. The high  temperature values are likely due to shocks produced by the outflows,  hence the gas reaches a maximum temperature in the immediate postshock region ($T_{\rm e} \propto V_{\rm s}^{2}$, \citealt{contini19}, where $V_{\rm s}$ is the shock velocity).  For Mrk\,79, it was possible to estimate $T_{\rm e}$ in the outskirt regions, where relatively low values  ($10\,000 - 12\,000$ K) are derived, which indicate a large temperature gradient in this object.  For Mrk\,348, the $T_{\rm e}$ maps show an increase in temperature from the nucleus to locations farther away from it, with  $T_{\rm e[N II]}$ increasing from $ 8\,000 $ to $\ga \:30\,000$ K. The  $T_{\rm e[O III]}$ map shows values larger than $\sim$20\,000 K in most locations, with slightly higher values seen in  regions away from the nucleus. 
The $T_{\rm e}$ distributions found for Mrk\,79 and Mrk\,348 are distinct from the findings of \citet{revalski18} for Mrk\,573 based on long slit data, where they find no systematic variations in temperature, with a mean value of  $13\: 500 \pm 650$ K. The $T_{\rm e[O III]}$ and $T_{\rm e[N II]}$ maps for Mrk\,607 show only small variations in temperature, with  $T_{\rm e[O III]}\sim$20\,000\,K and  $T_{\rm e[N II]}\sim$10\,000\,K. No clear evidence of outflows is seen in this galaxy \citep{Freitas18}.

Figure~\ref{golden} shows the  $T_{\rm e[N II]}$ versus  $T_{\rm e[O III]}$ plot with the observed values for  Mrk\,607 and Mrk\,348, which are compared to predictions of AGN photoionization
models built with the {\sc Cloudy} code \citep{ferland13} by \citet{dors20} and to integrated measurements for Seyfert 2 galaxies. While the points for Mrk\,607 are seen in the same region occupied by the  model predictions  and by  the integrated spectra, the points for Mrk\,348 are located in a very distinct position, towards higher temperatures. This result, together with the fact that the [\ion{O}{iii}] emission in the nuclear region of this galaxy is mainly due to outflowing gas \citep{Freitas18}, is a strong indication that outflows have an important effect
on temperature structure of AGN hosts.
We  find that the temperature increases with  [\ion{O}{i}]6300/H$\alpha$, a known tracer of shocks \citep[e.g.][]{allen08,rich14} and with the width of [\ion{O}{iii}]5007 (see Suppl. Mat.),  indicating that shocks from AGN winds produce the high temperatures observed in Mrk\,79 and Mrk\,348.

The spaxels which show the  largest values of $T_{\rm e}$ in our sample are seen approximately perpendicular to the AGN ionization axis (clearly observed in Mrk\,79, Fig.~\ref{fig:m79}), indicating that shocks have a larger relative contribution to the line emission at these locations. An increase in the [\ion{O}{iii}] line width is seen at these locations \citep[Sup. Mat.; ][]{Freitas18}, indicating the outflows are interacting with  gas clouds, partially ionizing them and producing the observed line emission. Photo-ionized gas is a more efficient emitter of [\ion{O}{iii}] lines than shock-ionized gas,  therefore, a possible shock contribution to the [\ion{O}{iii}] emission within the AGN ionization structure is overshadowed by the photoionization contribution. This result indicates that the outflows from Seyfert nuclei may have  wide-opening angles as observed in luminous AGN \citep[e.g.][]{kakkad20} and  predicted by theoretical models \citep{wagner12,ishibashi19}. Similar outflows in the equatorial plane of the torus have been reported in a few Seyfert galaxies \citep{rogemar_n5929,lena15}. The bipolar outflows observed in some cases \citep[e.g.][]{crenshaw10,sb_n4151} may be tracing only the photoionized part of the outflowing gas,  as shocks are not efficient in the line production, when competing with photoionization \citep{zakamska14}, and thus the outflows are observed within the AGN ionization structure.

\section{Conclusions}
\label{concsec}

We present  spatially resolved maps of the electron temperature within the inner few hundred parsecs of the Seyfert galaxies Mrk\,79, Mrk\,348, and Mrk\,607, obtained from integral field spectroscopy. Our results indicate that shocks play an important role in the observed electron temperature distributions, as they can not be reproduced by AGN photoionization models only. This provides a caveat to studies of chemical abundances based on the usual methods for objects with shock/outflows signatures. Not taking this properly into account could yield biased estimates of the gas properties because shocks can produce an electron energy distribution distinct from the Maxwell-Boltzmann  distribution. 
Shocks are more important in regions away from the AGN ionization axis, where they can be easily observed as the AGN radiation field is shielded by the nuclear dusty torus, while within the AGN ionization structure, photoionization is more efficient in producing line emission.

\section*{Acknowledgements}
We thank to an anonymous referee for the valuable comments that helped us to improve the paper.  
This study was financed in part by Conselho Nacional de Desenvolvimento Cient\'ifico e Tecnol\'ogico (202582/2018-3, 304927/2017-1 and 400352/2016-8), Funda\c c\~ao de Amparo \`a pesquisa do Estado do Rio Grande do Sul (17/2551-0001144-9 and 16/2551-0000251-7) and   do Estado de S\~ao Paulo. 
AF acknowledges the support from grant PRIN MIUR2017-20173ML3WW$_-$001.
EPM acknowledges financial support from  the project ``Estallidos6" AYA2016-79724-C4. 
Based on observations obtained at the Gemini Observatory, which is operated by the Association of Universities for Research in Astronomy, Inc., under a cooperative agreement with the NSF on behalf of the Gemini partnership: the National Science Foundation (United States), National Research Council (Canada), CONICYT (Chile), Ministerio de Ciencia, Tecnolog\'{i}a e Innovaci\'{o}n Productiva (Argentina), Minist\'{e}rio da Ci\^{e}ncia, Tecnologia e Inova\c{c}\~{a}o (Brazil), and Korea Astronomy and Space Science Institute (Republic of Korea). 
\section*{Data Availability}
The data used in this paper is available in the Gemini Science Archive under the project code GN-2014B-Q-87.


\bibliographystyle{mnras}
\bibliography{paper_arxiv} 



\appendix
\section{Examples of fits of the emission-line profiles}

The fitting of the spectra was performed using the {\sc ifscube} code \citep{ifscube}, as mentioned in Section~2. 
We perform a simultaneous fit of the following emission lines: H\,$\gamma$, [\ion{O}{iii}]$\lambda$\,4363, \ion{He}{ii}$\lambda$\,4686, H\,$\beta$, [\ion{O}{iii}]$\lambda\lambda$\,4959,5007, [\ion{Fe}{vii}]$\lambda$\,5721, [\ion{N}{ii}]$\lambda$\,5755, [\ion{O}{i}]$\lambda\lambda$\,6300,6364, [\ion{N}{ii}]$\lambda\lambda$\,6548,6583, H$\alpha$, and [\ion{S}{ii}]$\lambda\lambda$\,6716,6731. 
We allow the fit of up to three Gaussian components per emission line, tying the kinematics of the components of emission lines from the same parent ion, and keeping fixed the [\ion{N}{ii}]$\lambda$6583/$\lambda$6548 and [\ion{O}{iii}]$\lambda$5007/$\lambda$4959 flux ratios to their theoretical values of 3.06 and 2.98, respectively. We provide additional Gaussian components to account for the broad components of the H recombination lines in Mrk\,79, which hosts a type 1 AGN.  We fit the continuum emission by a fifth order polynomial. The code fits the nuclear spaxel using initial guesses provided by the user and than follows a spiral pattern from the nucleus outwards, using optimized guesses from the best-fit parameters from neighboring spaxels at distances smaller than 0\farcs35, by using the {\it refit} parameter of the {\sc ifscube} code.
 If the line profile is well reproduced by less than three Gaussian components at a specific spaxel, the amplitude of needless components are set to zero and the code returns the best-fit parameters of the remaining components.
We do not associate a physical meaning with each separate Gaussian component. The fluxes of each emission line is obtained by the sum of the fluxes of the individual components. The median flux uncertainties are 12\,\% for [\ion{O}{iii}]$\lambda$4363, and 15\,\% for  [\ion{N}{ii}]$\lambda$5755 but smaller than 5\,\% for the strong lines (e.g. [\ion{O}{iii}]$\lambda$5007).

\begin{figure*}
\includegraphics[width=0.49\textwidth]{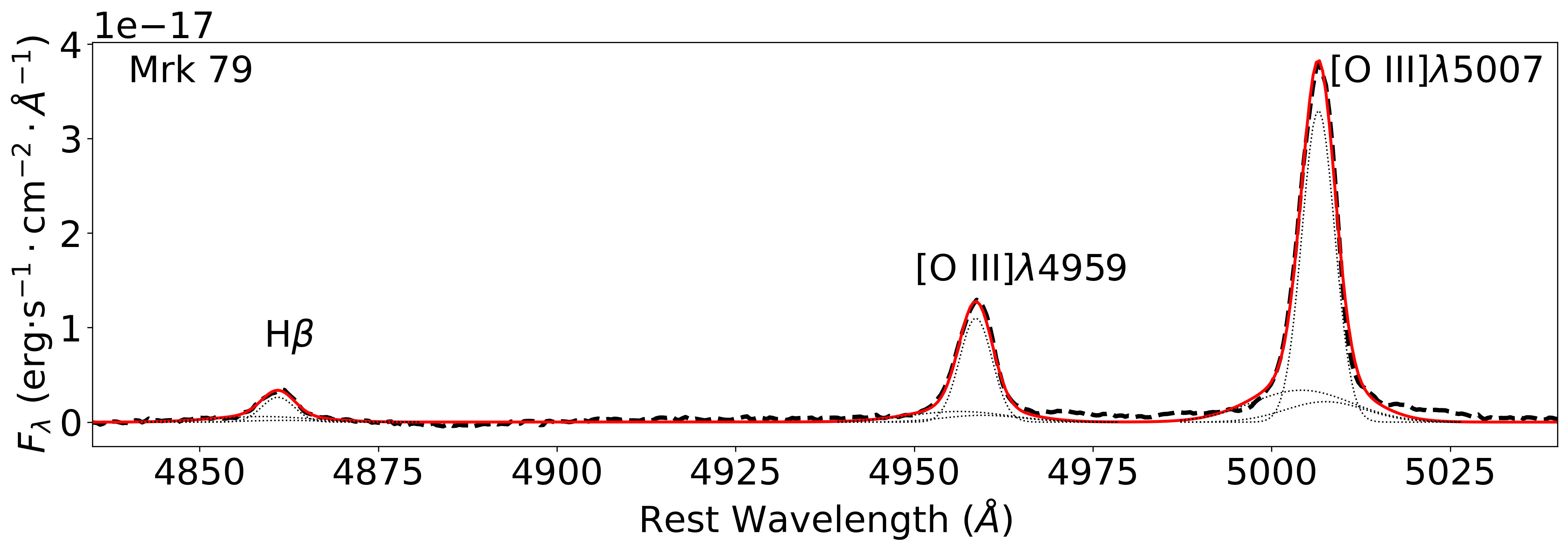}
\includegraphics[width=0.49\textwidth]{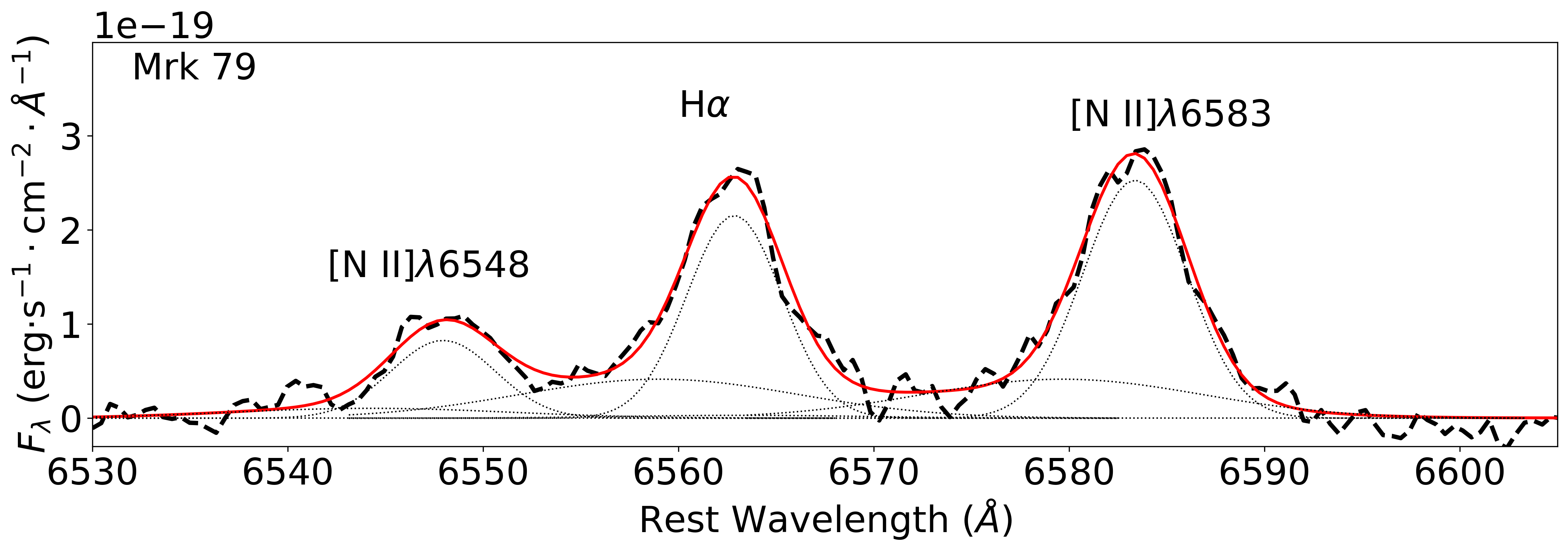}
\includegraphics[width=0.49\textwidth]{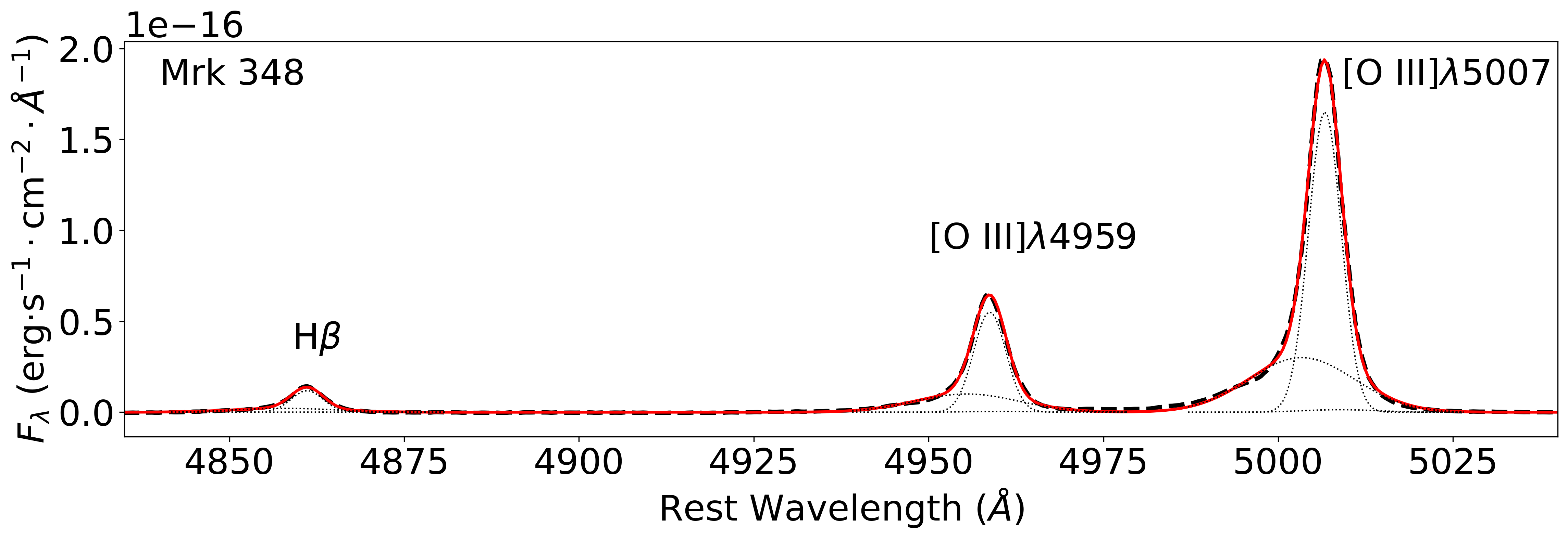}
\includegraphics[width=0.49\textwidth]{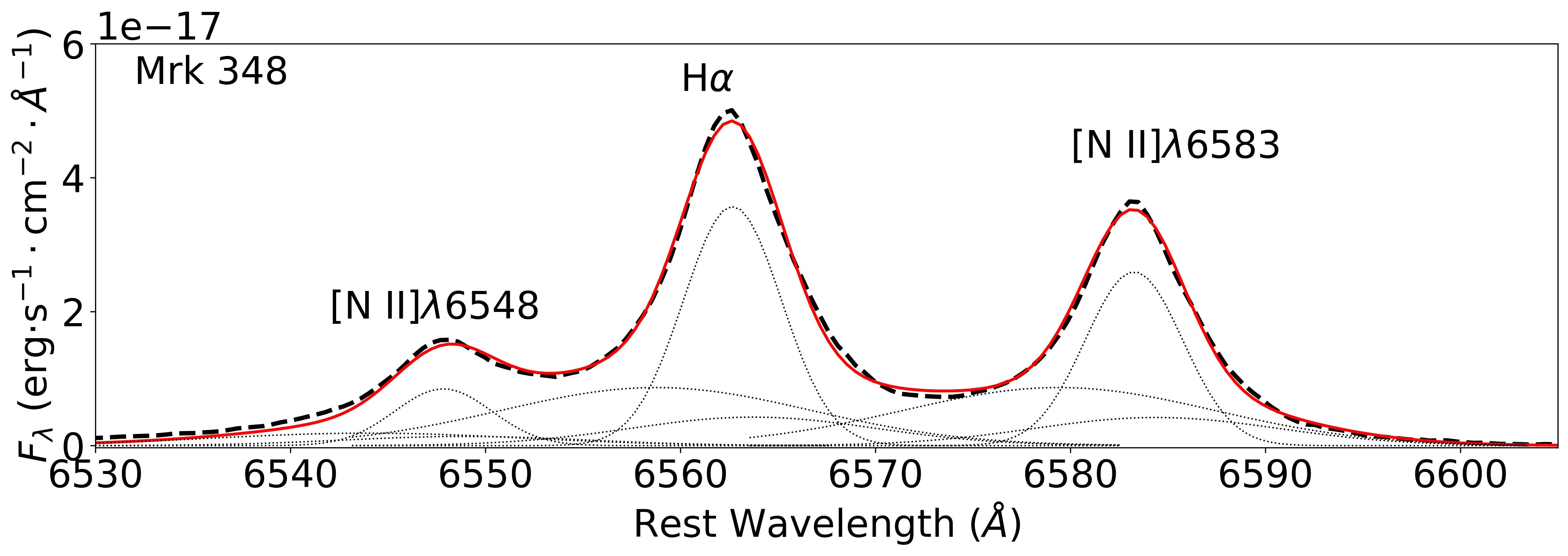}
\includegraphics[width=0.49\textwidth]{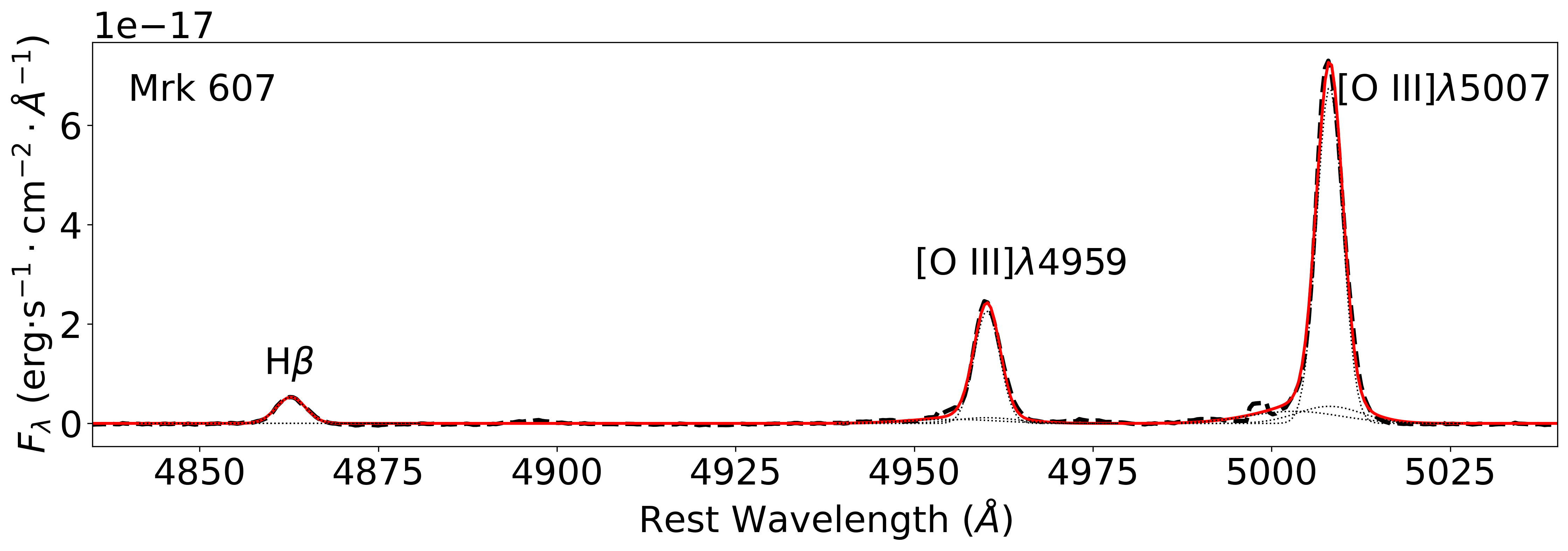}
\includegraphics[width=0.49\textwidth]{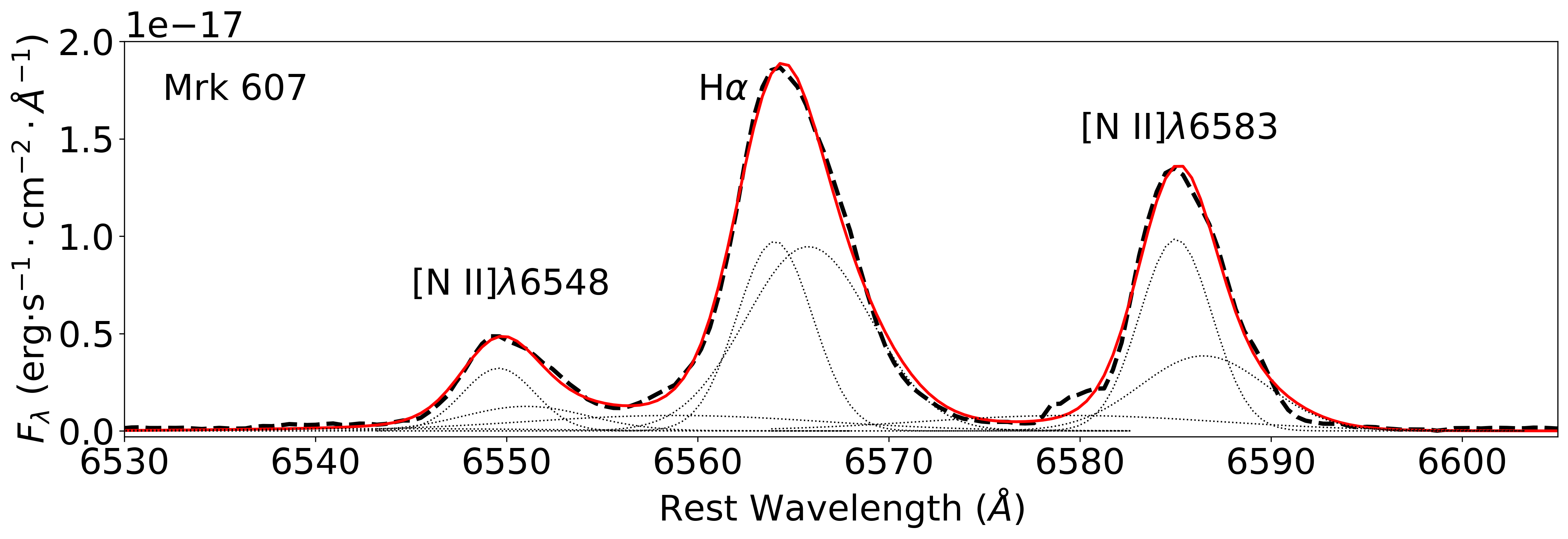}
\caption{ Examples of [\ion{O}{iii}]+H$\beta$ (left) and [\ion{N}{ii}]+H$\alpha$ (right) emission-line profiles for Mrk\,79 (top), Mrk\,348 (middle) and Mrk\,607 (bottom). The continuum-subtracted spectrum is shown in dashed black lines, the model in continuous red and the individual components are shown as dotted lines. For Mrk\,79, the contribution of the broad H$\beta$ and  H$\alpha$ emission has also been subtracted. } 
\label{fig:fits_strong}
\end{figure*}

\begin{figure*}
\includegraphics[width=0.33\textwidth]{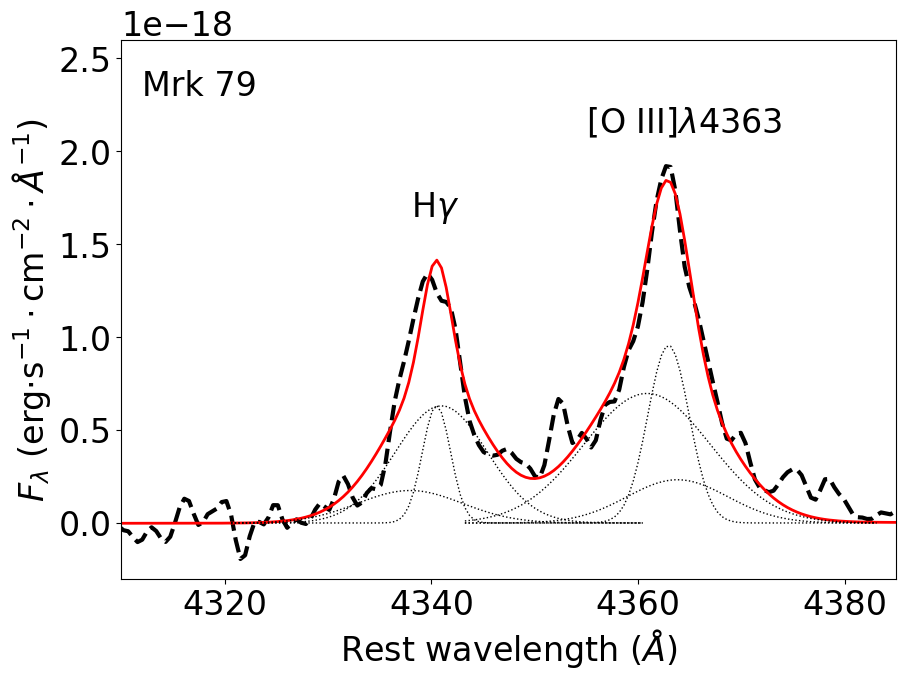}
\includegraphics[width=0.33\textwidth]{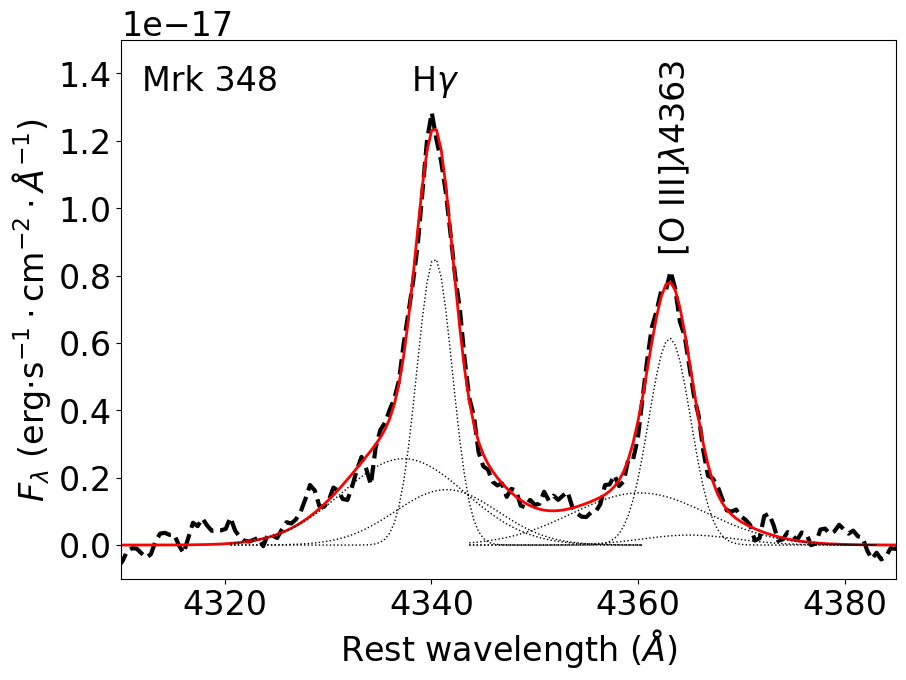}
\includegraphics[width=0.33\textwidth]{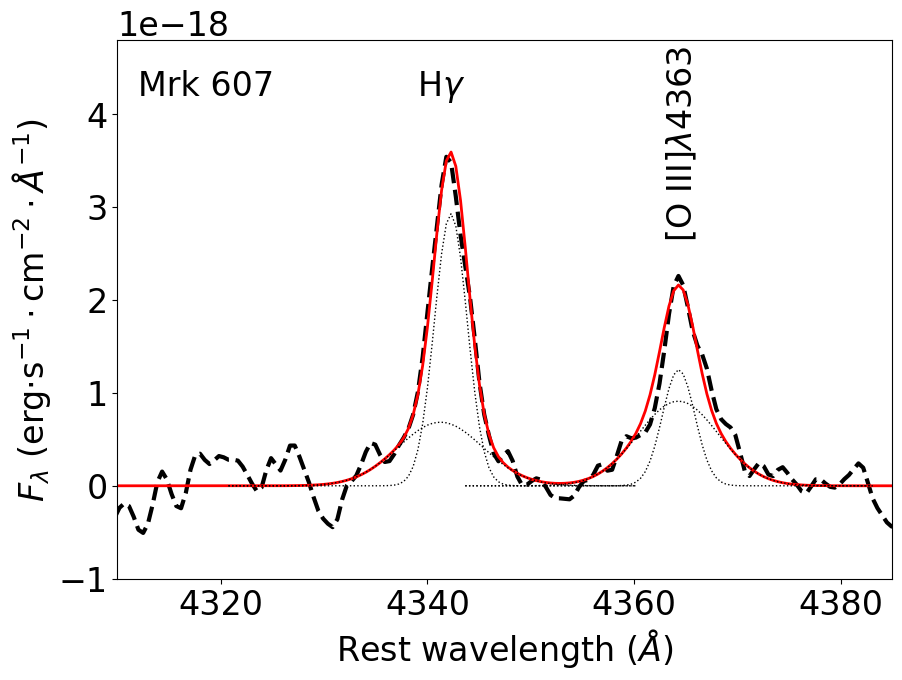}
\includegraphics[width=0.33\textwidth]{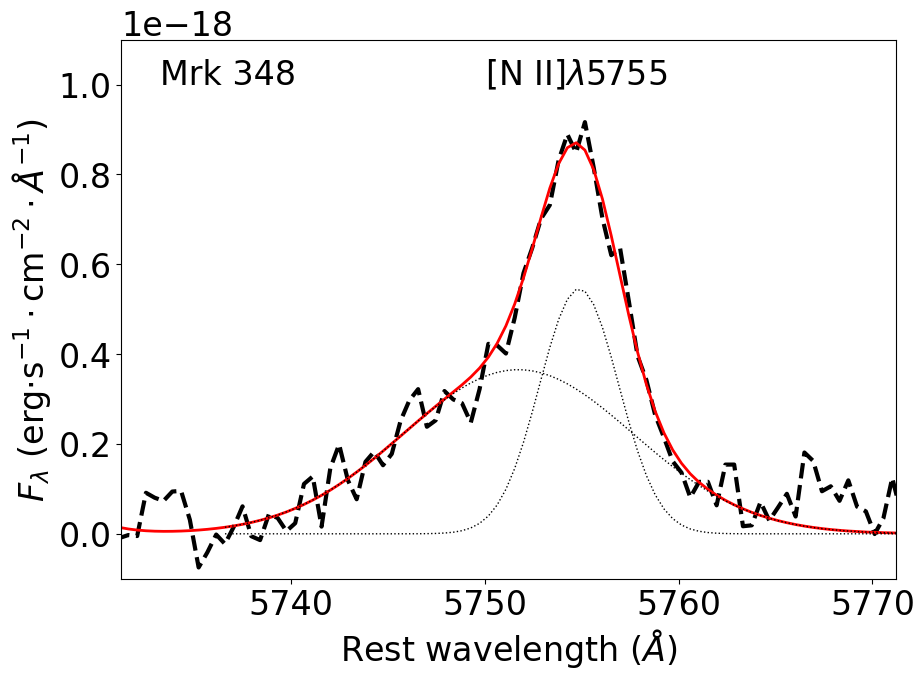}
\includegraphics[width=0.33\textwidth]{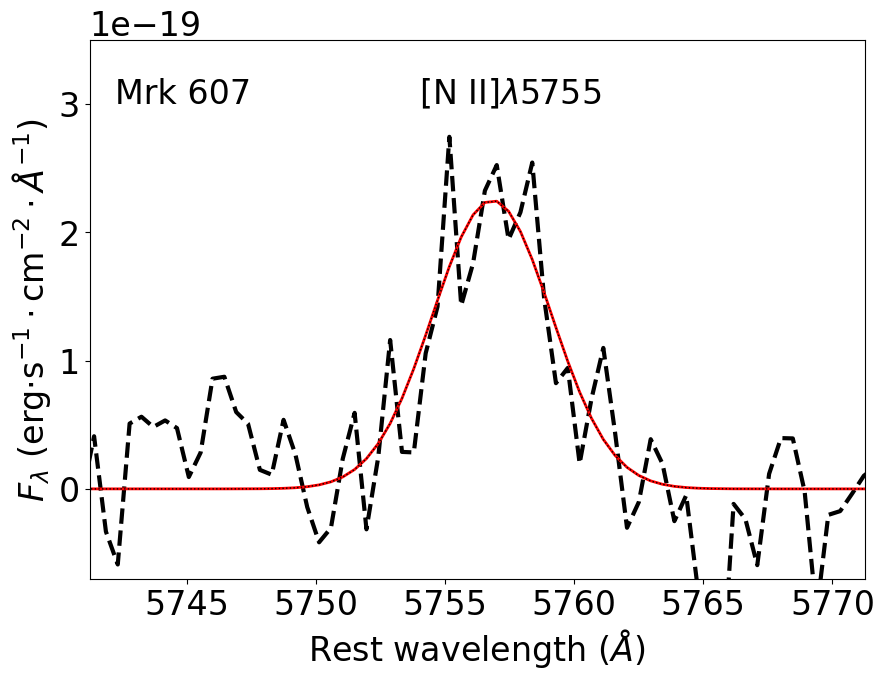}
\caption{Examples of the Gaussian fits of the H$\gamma$+[\ion{O}{iii}]$\lambda$4363 (top) and [\ion{N}{ii}]$\lambda$5755 (bottom). The continuum-subtracted spectrum is shown in dashed black lines, the model in continuous red and the individual components are shown as dotted lines. For Mrk\,79, the contribution of the broad H$\gamma$ emission has also been subtracted.}
\label{fig:fits}
\end{figure*}

Figure~\ref{fig:fits_strong} shows examples of the fits of the strong emission lines: H\,$\beta$, [\ion{O}{iii}]$\lambda\lambda$\,4959,5007,  H$\alpha$ and [\ion{N}{ii}]$\lambda\lambda$\,6548,6583. Figure~\ref{fig:fits} shows examples of the Gaussian fits of the H$\gamma$+[\ion{O}{iii}]$\lambda$4363 (top) and [\ion{N}{ii}]$\lambda$5755 (bottom) emission lines. The emission lines are well reproduced by the Gaussian models and the H$\gamma$ and [\ion{O}{iii}]$\lambda$4363 are clearly separated.

\section{Curves-of-growth for the [\ion{O}{iii}] and [\ion{N}{ii}] emission}
\label{apendCOG}

\begin{figure*}
\includegraphics[width=0.33\textwidth]{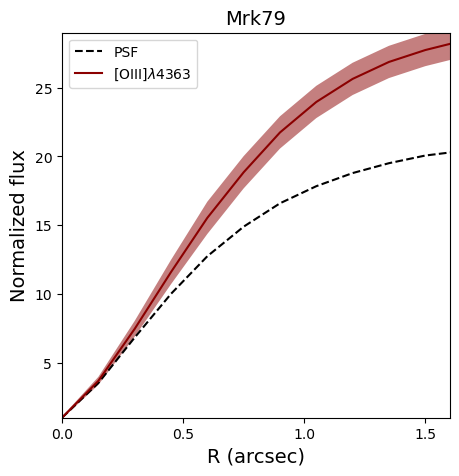}
\includegraphics[width=0.33\textwidth]{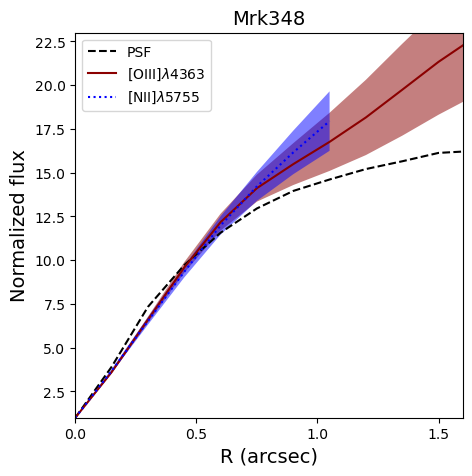}
\includegraphics[width=0.33\textwidth]{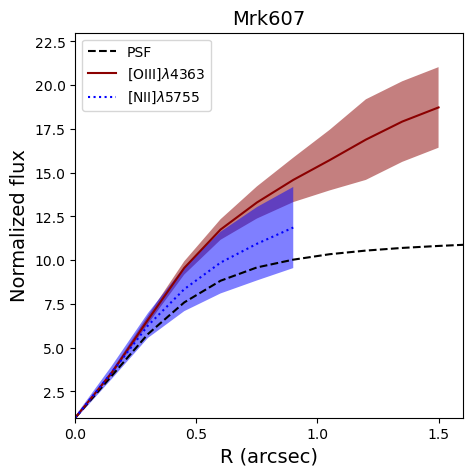}
\caption{Curves-of-growth for Mrk\,79 (left), Mrk\,348 (center) and Mrk\,607 (right). The black dashed line corresponds to the PSF as obtained from the flux distribution of the broad component of H$\alpha$ for Mrk\,79 and from field stars on the acquisition images of Mrk\,348 and Mrk\,607. The red and blue lines show the [\ion{O}{iii}]$\lambda$4363 and [\ion{N}{ii}]$\lambda$5755 curve-of-growth,  respectively. The shaded regions delineate the 1$\sigma$ flux uncertainties. The [\ion{N}{ii}]$\lambda$5755 is not detected for Mrk\,79. These curves were constructed by normalizing the flux distributions within the inner 0\farcs15 radius and computing the fluxes within apertures increasing in steps of 0\farcs15 each. }
\label{fig:psf}
\end{figure*}

The [\ion{O}{iii}]$\lambda$4363 and [\ion{N}{ii}]$\lambda$5755 flux distributions extend to distances larger than the FWHM of the point spread function (PSF) in all galaxies. However, the emission-line flux distributions of an unresolved nuclear source can be smeared by the seeing to distances larger than the FWHM of the PSF.  To further check if the [\ion{O}{iii}]$\lambda$4363 and [\ion{N}{ii}]$\lambda$5755 flux distributions in our sample are spatially resolved, we follow \citet{kakkad20} and compute their curves-of-growth (COGs). We compare the emission-line and PSF COGs, with the later obtained from images of field stars in the GMOS acquisition images for Mrk\,348 and Mrk\,607, and from the H$\alpha$ emission from the broad line region of Mrk\,79. First, we obtain the integrated flux within a circular aperture of 0\farcs15 centred at the nucleus, normalize the emission-line flux distribution by this value and then compute the integrated fluxes increasing the radius in steps of 0\farcs15. In Fig.~\ref{fig:psf} we show the corresponding COGs, which show the [\ion{O}{iii}]$\lambda$4363 emission presents an excess in flux as compared to the PSF, indicating that the [\ion{O}{iii}]$\lambda$4363 emission is spatially resolved in all galaxies. The [\ion{N}{ii}]$\lambda$5755 emission in Mrk\,348 is also spatially resolved, while in  Mrk\,607 it is marginally resolved.

\section{Emission-Line Ratio Maps}
\label{apendRatMaps}
The gas reddening is obtained from the H$_\alpha$ and H$_\beta$ observed fluxes by adopting the theoretical line ratio of ${I_{H\alpha}/I_{H\beta}}=2.86$ for Case B \ion{H}{i} recombination,  an electron temperature of $T_{\rm e} = 10\,000 $~K \citep{osterbrock06} and  $R_V = 3.1$. Using the  coefficients from the extinction law of \citet{cardelli89} the colour excess is given by:
\begin{equation}
    E(B-V) = 2.32~\log \left(\frac{(F_{\mathrm{H}\alpha}/F_{\mathrm{H}\beta})}{2.86} \right),
\end{equation}
where $F_{\mathrm{H}\alpha}$ and $F_{\mathrm{H}\beta}$ are the observed fluxes of the  ${\mathrm{H}\alpha}$ and ${\mathrm{H}\beta}$ emission lines at each spaxel.

\begin{figure*}
\includegraphics[width=0.98\textwidth]{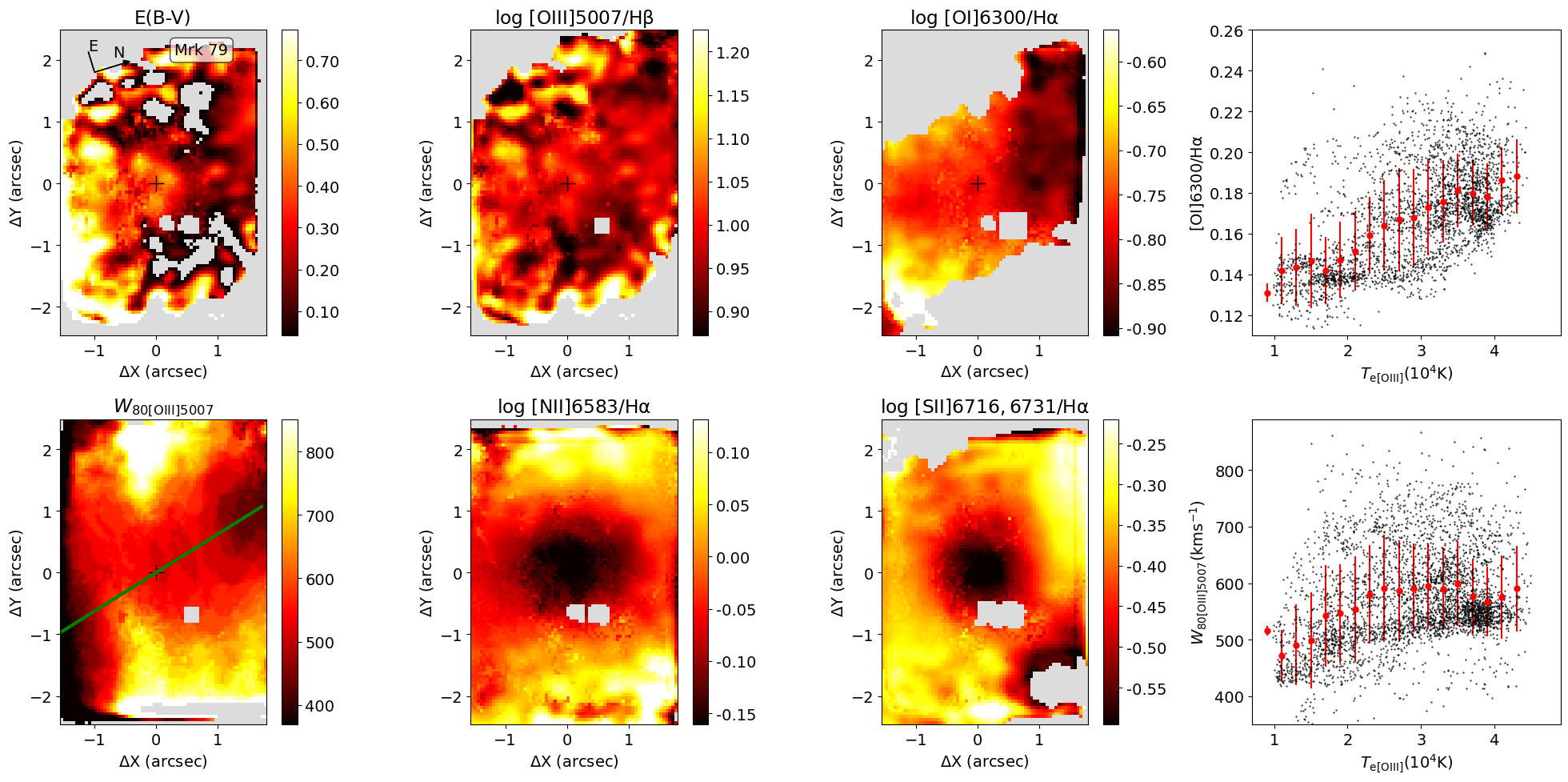}
\caption{Maps for Mrk\,79. Top row, from left to right: $E(B-V)$, log\, [\ion{O}{iii}]5007/H$\beta$ and log\,[\ion{O}{i}]6300/H$\alpha$ maps, and plot of $T_{\rm e[OIII]}$ vs. [\ion{O}{i}]6300/H$\alpha$. The black points show the observed  $T_{\rm e[OIII]}$ and [\ion{O}{i}]/H$\alpha$ values at each spaxel and the red circles show the mean values  within $T_{\rm e[OIII]}$ bins of $2\times10^{3}$ K and the error bars are the standard deviations of the [\ion{O}{i}]/H$\alpha$ within each bin.  Bottom row, from left to right: $W_{\rm 80}$ map for the [\ion{O}{iii}]5007 emission line -- the color bar show the  $W_{\rm 80}$ values in km\,s$^{-1}$, [\ion{N}{ii}]6583/H$\alpha$ and ([\ion{S}{ii}]6716+6731)/H$\alpha$ ratio maps, and $T_{\rm e[OIII]}$ vs. $W_{\rm 80[OIII]5007}$ plot. The points are defined in the same way as in the  $T_e$ vs. [\ion{O}{i}]/H$\alpha$ plot.  The green line shows the AGN ionization axis, as measured by \citet{schmitt03} using Hubble Space Telescope [\ion{O}{iii}]5007 narrow band images. }
\label{fig:bptsM79}
\end{figure*}

\begin{figure*}
\includegraphics[width=0.98\textwidth]{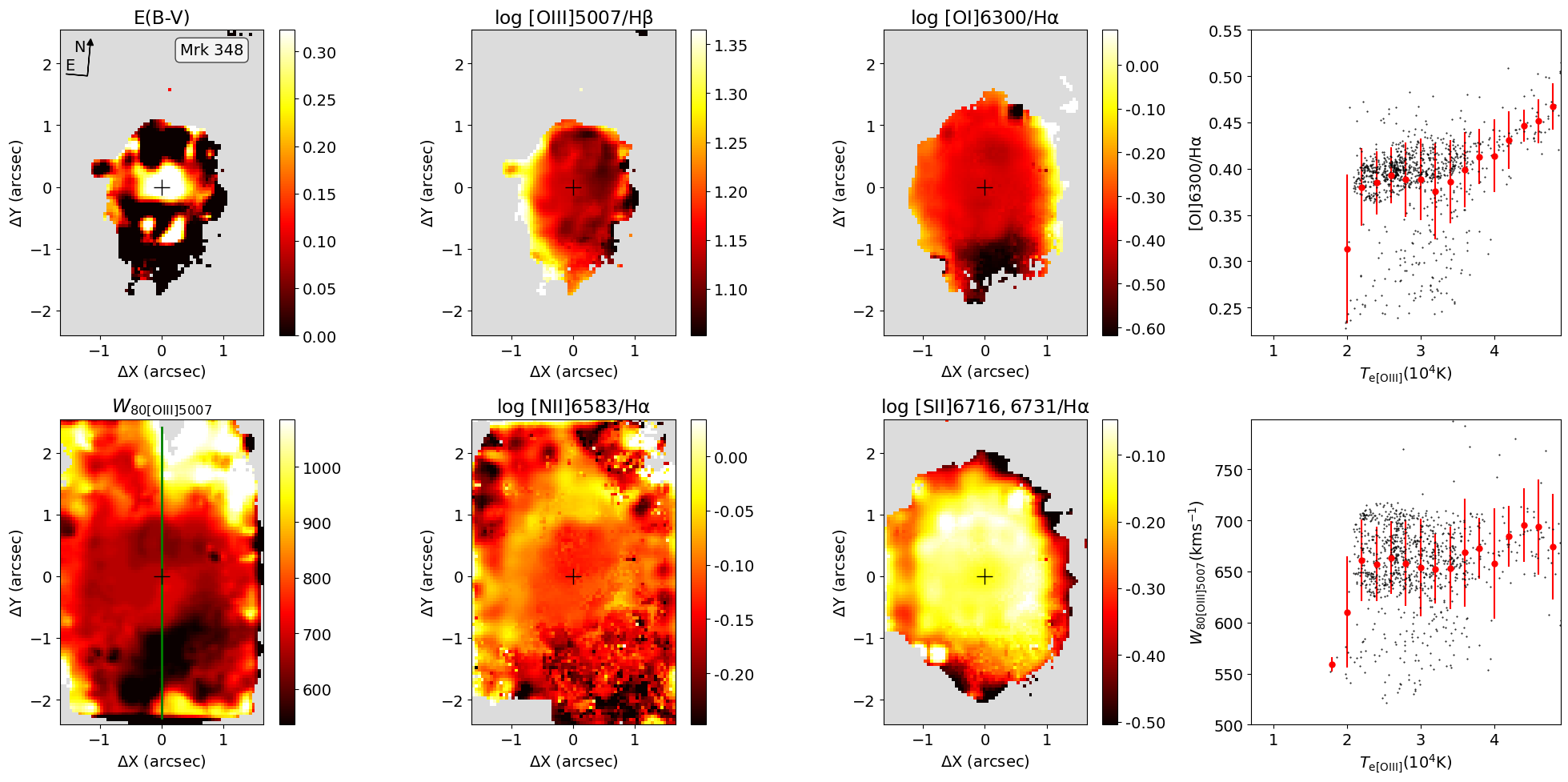}
\caption{Same as Fig.~\ref{fig:bptsM79}, but for Mrk\,348.}
\label{fig:bptsM348}
\end{figure*}

\begin{figure*}
\includegraphics[width=0.98\textwidth]{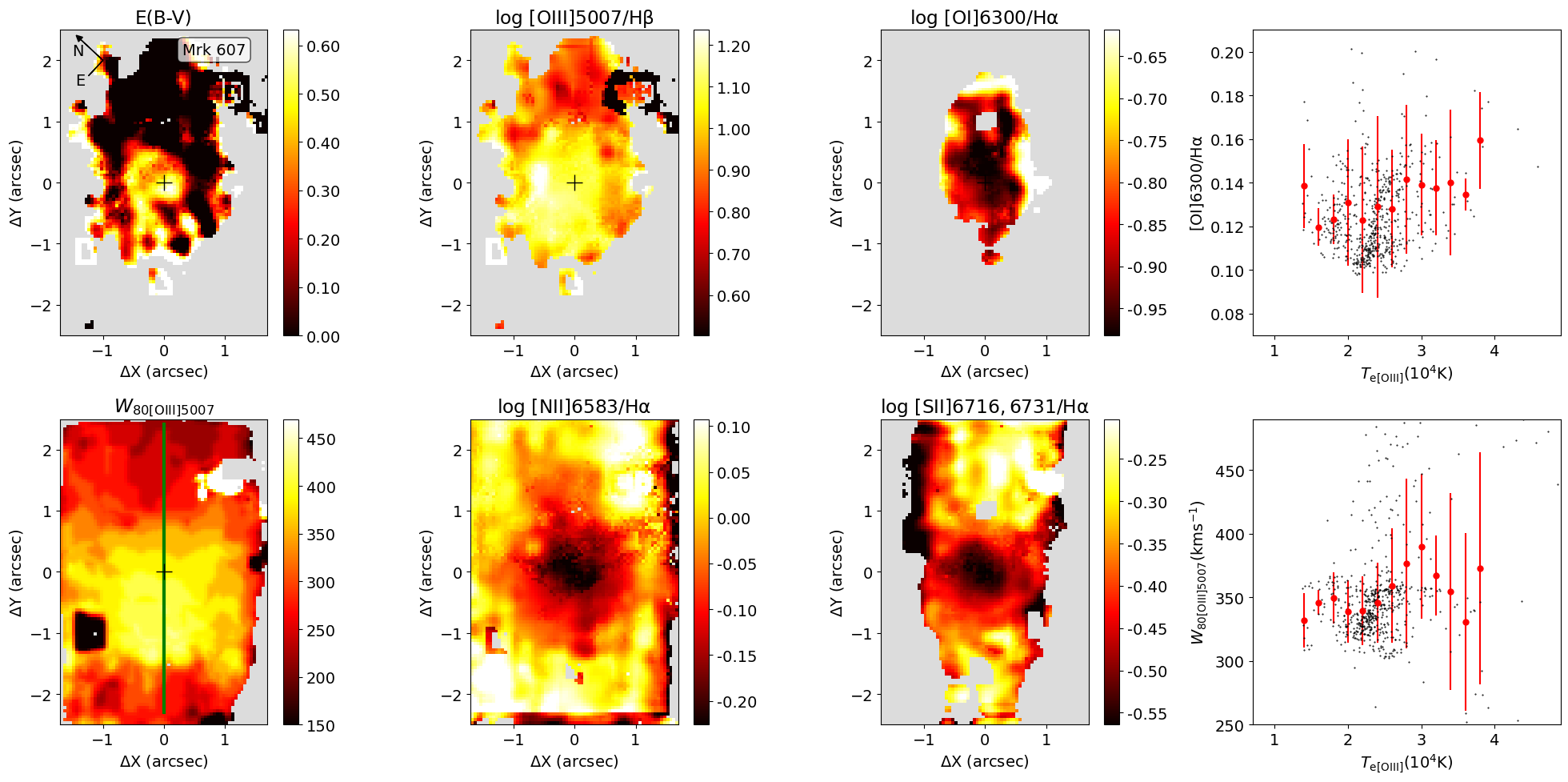}
\caption{Same as Fig.~\ref{fig:bptsM79}, but for Mrk\,607.}
\label{fig:bptsM607}
\end{figure*}

Figures~\ref{fig:bptsM79}, \ref{fig:bptsM348} and \ref{fig:bptsM607} show the $E(B-V)$ maps for Mrk\,79, Mrk\,348 and Mrk\,607, respectively. These figures also show  the [\ion{O}{iii}]5007/H$\beta$, [\ion{O}{i}]6300/H$\alpha$, [\ion{N}{ii}]6583/H$\alpha$ and ([\ion{S}{ii}]6716+6731)/H$\alpha$ flux ratio maps and $W_{\rm 80}$ maps for the [\ion{O}{iii}]5007 emission line. The flux ratio maps are corrected by extinction using the $E(B-V)$ values and adopting the extinction law by \citet{cardelli89}.  The $W_{\rm 80}$ parameter measures the width of the emission-line profile that includes 80\,\% of the integrated line flux. Values larger than 500-600~km\,s$^{-1}$ are commonly attributed to ionized gas outflows \citep{zakamska14,kakkad20,wylezalek20}.

\begin{figure*}
\includegraphics[width=0.32\textwidth]{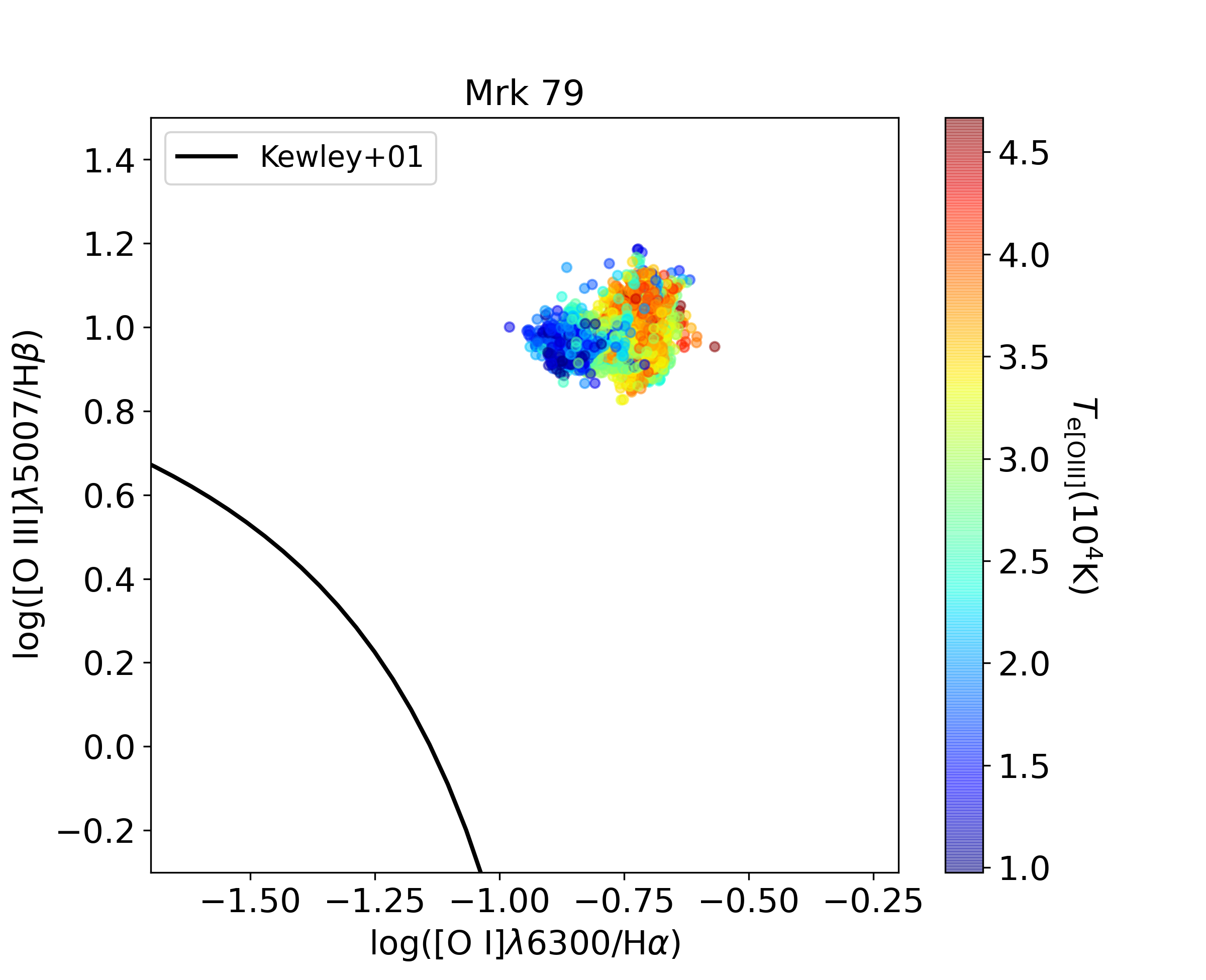}
\includegraphics[width=0.32\textwidth]{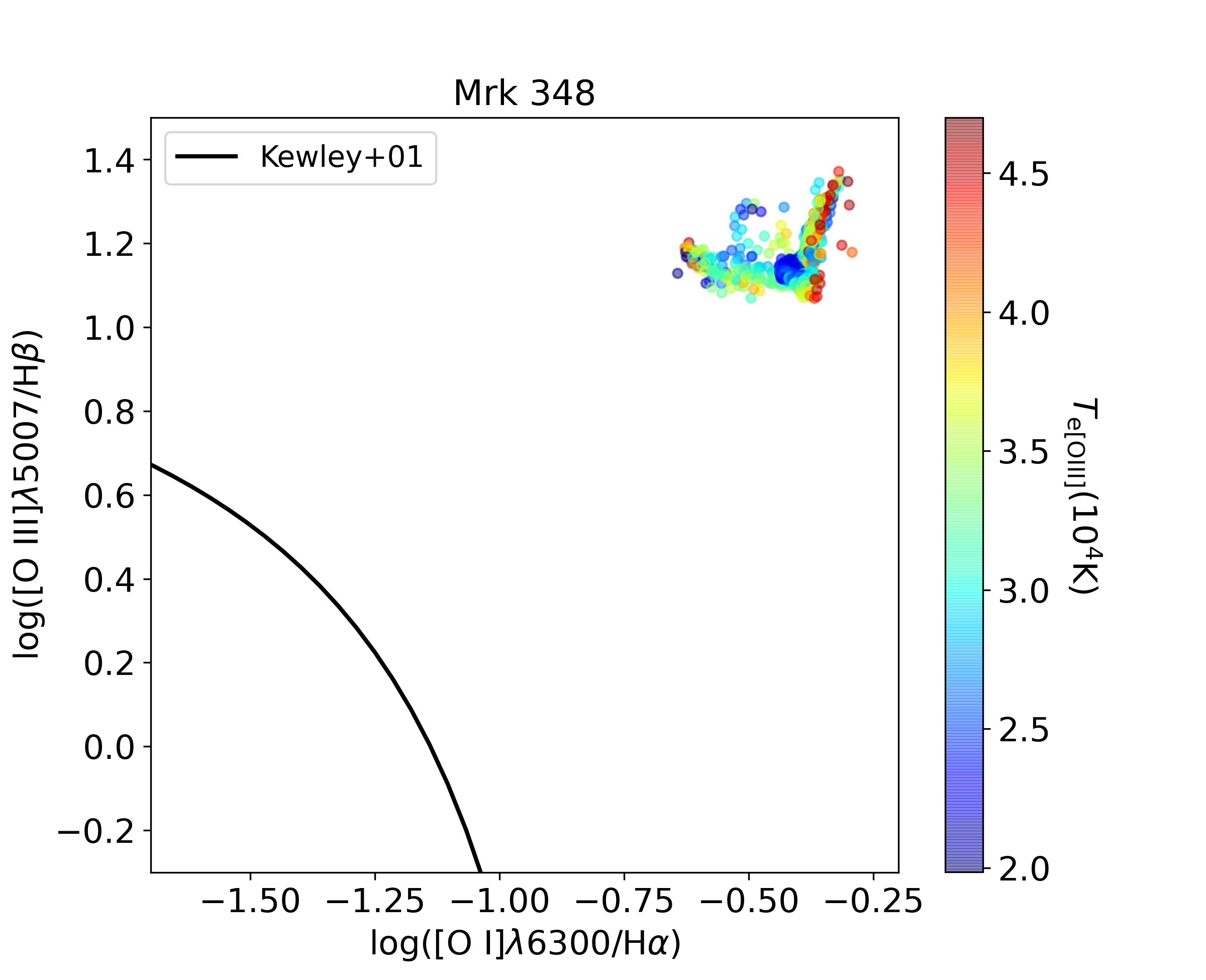}
\includegraphics[width=0.32\textwidth]{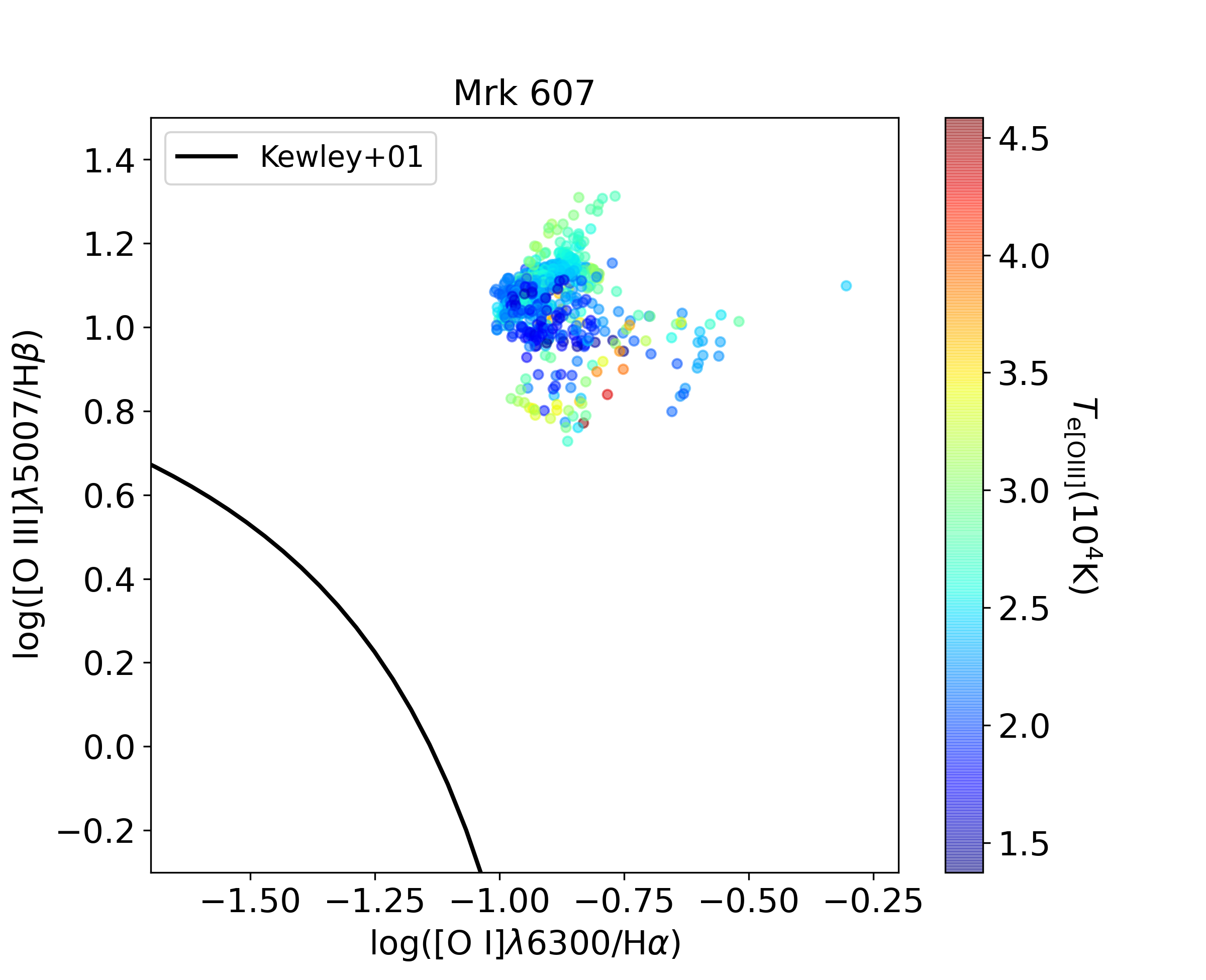}
\caption{BPT [\ion{O}{i}]-based diagrams for the spaxels with electron temperature measurements for Mrk\,79 (left), Mrk\,348 (middle) and Mrk\,607 (right). The continuous line is from \citet{kewley01}.}
\label{fig:bpts}
\end{figure*}

The flux-line ratios for all galaxies are typical of Seyfert nuclei, as already shown in \citet{Freitas18},  based on the fitting of the emission-line profiles by single Gaussian curves and using the same data considered here. Fig.~\ref{fig:bpts} shows the [\ion{O}{i}]-BPT diagram \citep{bpt89} for the spaxels with $T_{\rm e[OIII]}$ estimates. The points are color coded using by the $T_{\rm e[OIII]}$ values of each spaxel, as shown in the color bar.

The [O\,{\sc i}]$\lambda$6300/H$\alpha$ line ratio is a tracer of shocks in neutral gas \citep[e.g., ][]{allen08,rich11,rich14}. If the velocity dispersion of [O\,{\sc i}]  is larger than 150 km\,s$^{-1}$ and [O\,{\sc i}]$\lambda6300$/H$\alpha \gtrsim0.1$, shocks with velocities in the range of 160--300\, km\,s$^{-1}$ are the dominant excitation mechanism of the [O\,{\sc i}] emission line. The upper right panels of Figs.~\ref{fig:bptsM79}, \ref{fig:bptsM348} and \ref{fig:bptsM607} show a plot of [O\,{\sc i}]$\lambda6300$/H$\alpha$ vs. $T_{\rm e[OIII]}$ for the galaxies of our sample. For Mrk\,79 and Mrk\,348, the mean [O\,{\sc i}]$\lambda6300$/H$\alpha$ ratio increases with increasing temperature, reaching individual ratio values of up to 0.25 and 0.50 for Mrk\,79 and Mrk\,348. Mrk\,607 presents overall smaller values of [O\,{\sc i}]$\lambda6300$/H$\alpha$ and there is no trend of this ratio with the electron temperature. This result indicates that shocks play an important role in the production of the highest electron temperatures in Mrk\,79 and Mrk\,348. The observed  [O\,{\sc i}]$\lambda6300$/H$\alpha$ in Mrk\,79 and Mrk\,348 can be produced by shocks with velocities of $V_S \gtrsim300$ km\,s$^{-1}$ \citep{allen08}, which are smaller than the observed outflow velocities traced by the $W_{\rm 80}$ parameter.

The shocks in Mrk\,79 and Mrk\,348 may be produced by outflows. Both galaxies show well defined bipolar outflows along the AGN ionization axis \citep{rogemar_mrk79,Freitas18}. In addition, the high  $W_{\rm 80}$ values for the [\ion{O}{iii}]5007 line in these galaxies (bottom-left panels of Figs.~\ref{fig:bptsM79} and \ref{fig:bptsM348}) indicate that the presence of outflows not only along the ionization axis, but also in locations away from it. Values of $W_{80}\gtrsim500$ km\,s$^{-1}$ are commonly interpreted as tracers of AGN outflows \citep{zakamska14,kakkad20,wylezalek20}. 
For Mrk\,79, the highest $W_{\rm 80}$ values are observed perpendicular to the AGN ionization axis, which is indicated as a continuous line in the $W_{\rm 80}$ map of Fig.~\ref{fig:bptsM79}. This galaxy shows radio emission approximately along the same direction of the AGN  axis \citep{schmitt01}, while Mrk\,348 and Mrk\,607 do not present signature of radio jets \citep{nagar99}. The high $W_{\rm 80}$ values in regions away from the AGN ionization cone indicates that the AGN winds are more spherical and not restricted to the ionization cone, in agreement with observations of luminous AGNs \citep[e.g.][]{kakkad20} and  predicted by theoretical models \citep{ishibashi19}. Similar outflows approximately in the equatorial plane of the torus have been observed in a few Seyfert galaxies using optical and near-IR IFS \citep{rogemar_n5929,lena15}. The interaction of these outflows with the ambient gas may produce shocks, responsible for the increase of the temperature in regions away from the AGN ionization cone.  

\bsp	
\label{lastpage}
\end{document}